# First-principles calculations of thermal expansion path of the $Co_7Mo_6$ μ-phase

**Dmitry Vasilyev**

Baikov Institute of Metallurgy and Materials Science of RAS, 119334, Moscow, Leninsky Prospekt 49, Russia

dvasilyev@imet.ac.ru ; vasilyev-d@yandex.ru

**Abstract**.

The μ-phase, which is an intermetallic compound, can form in steels and superalloys, and significantly degrade their mechanical properties. The thermal expansion of $Co_7Mo_6$ μ-phase have been investigated by first-principles based quasi-harmonic Debye - Grüneisen approximation and by comparing the free energies calculated along different paths of thermal expansion. The calculations were carried out for temperatures from 0 K to 1500 K which below the decomposition temperature of $Co_7Mo_6$. The electronic, vibrational and magnetic energy contributions to the free energy were accounted. The thermal expansion of $Co_7Mo_6$ is not isotropic. The influence of physical factors on the stability of $Co_7Mo_6$ was analyzed. The isobaric heat capacity, thermal expansion, elastic constants, bulk modulus, sound wave velocities, Debye and Curie temperatures were reported. The calculated results analyzed and are in an agreement with the available experimental and theoretical data.

*Keywords:* Mu-phase; First-principles calculations; Thermal expansion; Elastic properties; Debye temperature; Thermodynamic properties

## 1. Introduction

The $Co_7Mo_6$ compound is the topologically close-packed (TCP) intermetallic phase existing in the Co-Mo systems. In the Co-Mo phase diagram the μ-phase has a homogeneity range from approximately 45 to 50 at. % Mo as presented by Davydov et al. [1, 2]. Thus, the $Co_7Mo_6$ with its stoichiometry composition of 46.2 at. % of Mo is a stable compound.

The $Co_7Mo_6$ compound is the model μ-phase which used to study the effect of precipitation of TCP phases on mechanical properties of the cubic matrices of cobalt (or nickel) based superalloys, because it is prone to precipitation during thermal exposure at high temperature. A large number of refractory elements have been adding to superalloys to improve their creep and stability at elevated temperatures, but the drawback of it is a precipitation of TCP phases [3].

The effect of a possible precipitation of TCP phases brings significant obstacles in design of superalloys which are widely used in production of turbine blades for aircraft engines. It's because, these materials have excellent high-temperature mechanical properties, creep and corrosion resistance [4].

The μ-phase is a hard-brittle compound and is detrimental because its precipitation induces a local depletion of a matrix of superalloys and its needle morphology can easily initiate a rupture and influence in particular the creep resistance, yield strength and ductility. Therefore, the range of stability of μ-phase should be accurately investigated. Especially, Gibbs potentials, thermodynamic properties and a path of thermal expansion of μ-phase must be calculated to model phase boundaries and simulate properties of superalloys.

The alloy development depends on existing phase diagrams and reliable thermodynamic description of Gibbs energies of constituent phases. Due to slow diffusion and the presence of other phases, it can be difficult to conduct an experiment at low temperatures, lower than 1073 K. Therefore, quantum mechanical calculations together with models can significantly help in obtaining such a thermodynamic description of the concerning phases.

In this work, the calculations are restricted to $Co_7Mo_6$ compound and does not take into account the disorder associated with possible deviation from the stoichiometry.

The μ-phase due to its technological importance attracts a lot of attention. A number of works studding the phase equilibria, structural properties and lattice parameters of μ-phases were conducted over the last years. The phase boundaries of the μ-phase in the Co-Mo system were studied in the work [1, 2]. The experimental calorimetric study of the $Co_7Mo_6$ was reported by Spencer et al. [5]. The phase equilibria in the $Co_7Mo_6$ μ-phase were studied by Oikawa et al. [6]. Zhao et al. [7] studied a precipitation of μ-phase in nickel base alloys by scanning, transmission electron microscopy and X-ray diffraction. Qin et al. [8] investigated the μ-phase in superalloys to applied stress at T = 800 C. The site occupation in the Ni-Nb μ-phase and configurational thermodynamic properties have been studied by Sluiter et al. [9]. The formation enthalpies of elements of Fe and Mo in metastable μ-phases were reported by Sluiter in [10], using the DFT calculations. Joubert et al. [11] studied the atom distribution on the five different sites of the crystal structure of μ-phases. The structure and lattice parameters of $Co_7Mo_6$ μ-phase have been studied by Forsyth et al. [12].

As one could see from above, that these studies are mainly focusing on structure properties of μ-phases. But the thermodynamic and physical properties of $Co_7Mo_6$ compound are not reported yet. The factors influencing on stability of μ-phase have not been investigating so far. But, Grabowski et al. [13] have shown that the magnetic entropy can significantly increase, up to 30 %, the free energy of alloys and it should be accounted in calculations. Moreover, the answer to the question of whether the thermal expansion of these compounds is isotropic or not is also unknown.

Insofar as answering these questions and obtaining these data are important for calculating the Gibbs potentials and predicting the properties of compounds at various temperatures, this was the motivation for this study.

In this work, the thermal expansion of the $Co_7Mo_6$ μ-phase was studied by first-principles calculations based on density functional theory (DFT) and combined with the quasi-harmonic Debye–Grüneisen approximation (QHA), as used in [13, 14]. Since, the crystal lattice of $Co_7Mo_6$ has two parameters *a* and *c*, and in order to find out how the lattice parameters of $Co_7Mo_6$ change during heating up, the approach used to calculate thermodynamic properties of Laves phase $Fe_2Mo$ in the work [15], was utilized.

The electronic and the vibration energies, the magnetic entropy of local magnetic moments of atoms were accounted to study the thermodynamic properties and evaluate their effect on the stability of the $Co_7Mo_6$. The thermal expansion path, isobaric heat capacity, structural properties, elastic constants of the strain tensor, bulk, shear, Young's modulus, Debye and Curie temperatures are calculated. The anisotropic thermal expansion in the $Co_7Mo_6$ compound is predicted. The results of this work can be useful for further modelling the Gibbs energies of the μ-phase as a function of composition, for calculating the phase diagrams and further theoretical and experimental research. In addition, it may provide valuable data which difficult to obtain from experiments.

## 2. Method of calculation

### *2.1. Thermodynamic model*

Helmholtz free energy *F(V,T)* was formulated according to [13, 16]

$$F(V,T) = E_{tot}(V) + F_{el}(V,T) + F_{vib}(V,T) + F_{mag}(V,T) - TS_{conf} \qquad (1)$$

where $E_{tot}(V)$ is the total energy obtained by DFT calculations at T = 0 K. The other free energy terms are the electronic, $F_{el}(V,T)$; vibrational, $F_{vib}(V,T)$; magnetic, $F_{mag}(V,T)$ and ideal configurational entropy, $S_{conf}$.

### *2.2. Free energy subsystems*

The electronic energy $F_{el}(V,T)$ was described as in [17]

$$F_{el}(V,T) = E_{el}(V,T) - TS_{el}(V,T), \qquad (2)$$

With the energy of electrons due to their excitations $E_{el}(V,T)$ is formulated by [18]

$$E_{el}(V,T) = \int_{-\infty}^{\infty} n(\varepsilon,V) f(\varepsilon,T)\varepsilon d\varepsilon - \int_{-\infty}^{\varepsilon_F} n(\varepsilon,V)\varepsilon d\varepsilon \qquad (3)$$

where $n(\varepsilon,V)$ is the electronic density of states DOS, $f(\varepsilon,T)$ is the Fermi-Dirac distribution. The electronic entropy $S_{el}(V,T)$ takes the form

$$S_{el}(V,T) = -k_B \int_{-\infty}^{\infty} n(\varepsilon,V)\big(f(\varepsilon,T)\, lnf(\varepsilon,T) + \big(1 - f(\varepsilon,T)\big) ln(1 - f(\varepsilon,T))\big) d\varepsilon \qquad (4)$$

where $k_B$ is the Boltzmann constant.

The vibrational free energy of the lattice ions under the quasi-harmonic Debye- Grüneisen approximation it is given by [13, 14]

$$F_{vib}(V,T) = E_D(V,T) - TS_{vib}(V,T), \qquad (5)$$

where the energy of ions due to their vibrations $E_D(V,T)$ is formulated by

$$E_D(T,V) = \frac{9}{8} N_A k_B \theta_D + 3N_A k_B T D\left(\frac{\theta_D}{T}\right) \qquad (6)$$

$$S_{vib}(T,V) = 3N_A k_B \left[\frac{4}{3} D\left(\frac{\theta_D}{T}\right) - ln(1 - exp\left(-\frac{\theta_D}{T}\right))\right] \qquad (7)$$

Where, D($\theta_D$/T) – Debye function; The Grüneisen parameter γ was expressed according to [14]

$$\gamma = -1 - \frac{V}{2} \frac{\partial^2 P/\partial V^2}{\partial P/\partial V} \qquad (8)$$

The Debye temperature $\theta_D(V)$ was formulated as [14]

$$\theta_D(V) = \theta_{Do}\left(\frac{V_0}{V}\right)^{\gamma} \qquad (9)$$

where $\theta_{D0}$ - the Debye temperature obtained at T = 0 K and equilibrium volume, $V_0$, which are listed in Table 4 and Table 1 respectively.

The magnetic energy $F_{mag}(V,T)$ was described by the assumption:

$$F_{mag}(V,T) = [F'_{mag}(T) - F'_{mag}(0K)] - TS_{mag}(V) \qquad (10)$$

Where $F'_{mag}(T)$ is the magnetic energy term of Hillert and Jarl model as given in [19]

$$F'_{mag}(T) = RTLn(\beta + 1) f(\tau) \qquad (11)$$

where $\tau = T/T_c$; the function $f(\tau)$ is defined in [19]; $\beta$ is the average magnetic moment of atoms. The expression in square brackets of (10) is to compensate the constant contribution of magnetic internal energy persistently existing in the total energy.

The theoretical estimation of Curie temperature was calculated using the mean-field approximation borrowing in [13]

$$T_C = \frac{2\left(E_{tot}^{PM}(V_0) - E_{tot}^{FM}(V_0)\right)}{3k_B} \quad (12)$$

where $E^{PM}{}_{tot}(V_0)$ and $E^{FM}{}_{tot}(V_0)$ are the total energies of the compound calculated at T = 0 K for paramagnetic (PM) and ferromagnetic (FM) states at the corresponding equilibrium volumes $V_0$.

The magnetic entropy $S_{mag}(V)$ and configurational entropy $S_{conf}$ were calculated as follow:

$$S_{mag}(V) = k_B \sum_{i=1}^{n} c_i ln\,(|\mu_i(V)| + 1) \quad (13)$$

$$S_{conf} = k_B \sum_{i=1}^{n} c_i ln\, c_i \quad (14)$$

where $\mu_i$ is the local magnetic moment of atom $i$, $c_i$ is the atomic concentration.

### *2.3. First-principles calculations*

The calculations were carried out using the Full Potential-Linear Augmented Plane Wave (FP-LAPW) Method [20] within the (DFT) [21, 22], as implemented in the WIEN2k package [20]. The exchange-correlation effects were treated as a function within the Generalized Gradient Approximation (GGA) [23] and GGA along with Perdew-Bruke-Eruzerhof (PBE) [24]. The muffin-tin radius ($R_{MT}$) were set as $R_{MT}$ = 2.05 for Co, $R_{MT}$ = 2.15 for Mo. The parameter $RK_{max}$ was set as 8. To model the $Co_7Mo_6$ compound the two kinds of cells, in hexagonal [11] and rhombohedral [12] notation of 39 and 13 atoms, were used with the 15 x 15 x 3 and 12 x 12 x 12 $k$-point mesh in the first irreducible Brillouin zone using Monkhorst-Pack scheme [25]. The energy between successive iterations is converged to $1\times10^{-6}$ eV/atom and forces are minimized to 5 meV/Å.

## 3. Results and discussion.

### *3.1. Ground state properties*

#### *3.1.1. Structural properties*

The crystal lattice of $Co_7Mo_6$ μ-phase belongs to the *R3m* space group and typified by $Fe_7W_6$ compound. The unit cell of $Co_7Mo_6$ can be represented in rhombohedral and hexagonal notations, with 13 and 39 atoms per unit cell, according [12, 11]. The hexagonal notation of $Co_7Mo_6$ is shown in Figure 1.

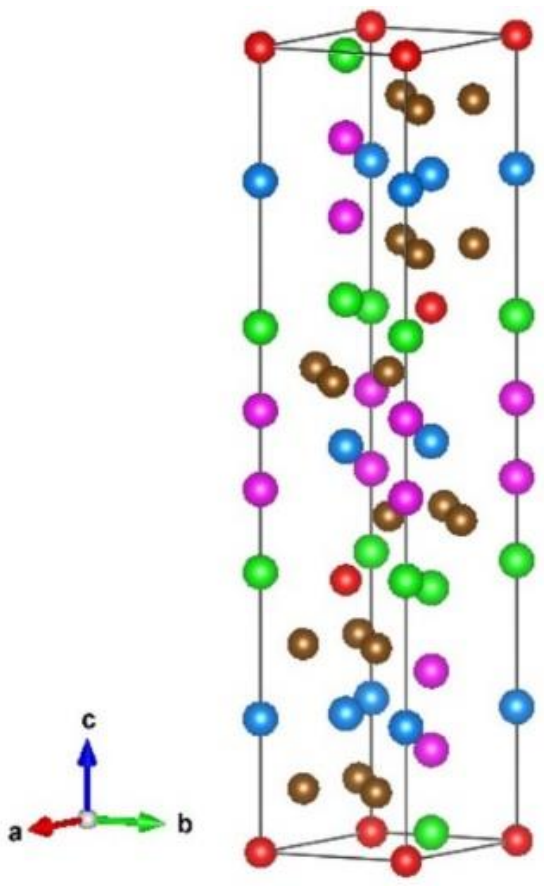

**Figure 1**. Unit cell of $Co_7Mo_6$ μ-phase, in the hexagonal notation. Co atoms occupy *3a* and *18h* Wyckoff sites are shown in red and brown respectively while Mo atoms occupy *6c*, *6c'* and 6*c"* sites are shown in blue, green and pink respectively.

The computations in this work were carried out with spin-polarized calculations. The geometry optimizations and the full structural relaxation procedures were used to obtain lattice parameters and atoms positions of $Co_7Mo_6$ compound. The optimized lattice parameters are listed in Table 1 and Figure 3, together with the available experimental data [12, 26]. The lattice parameters of $Co_7Mo_6$ calculated in this work at T = 1423 K are close to the experimental data [26] obtained at T = 1423 K after 912 hours of annealing treatments, as listed in Table 1 and shown in Figure 3. The enthalpy formation $\Delta H$ of $Co_7Mo_6$ was calculated by the following expression

$$\Delta H^{Co7Mo6} = E_{total}^{Co7Mo6} - \left(xE_{bcc}^{Mo} + (1-x)E_{hcp}^{Co}\right) \qquad (15)$$

where $E^{Co7Mo6}_{total}$ is the total energy of $Co_7Mo_6$, $E^{Co}_{hcp}$ and $E^{Mo}_{bcc}$ are the energies of Co and Mo atoms in the hexagonal close-packed (hcp) and the body centred cubic (bcc) lattice, respectively; $x$ is the concentration of Mo atoms ($x$ = 0.462). The $\Delta H$ of $Co_7Mo_6$ calculated in this work have the negative value, this suggests that the compound is stable at T = 0K, the obtained value is listed in Table 1 with the other theoretical value [27] for comparison.

**Table 1**
Ground state properties of lattice constants (in Å), equilibrium volume V (in Å$^3$), formation enthalpies ΔH (kJ/mol) of $Co_7Mo_6$ together with experimental and other theoretical values.

| Compound | Method | *a* | *c* | *c/a* | *V* | Δ*H* |
|---|---|---|---|---|---|---|
| $Co_7Mo_6$ | This work | 4.744 | 25.423 | 5.359 | 495.524 | -4.43 |
| | This work [1)] | 4.756 | 25.644 | 5.392 | 502.388 | |
| | exp. [26] [2)] | 4.761 | 25.59 | 5.375 | 502.34 | |
| | exp. [12] | 4.762 | 25.615 | 5.379 | 503.04 | |
| | calc. [27] | | | | | -2.51 |

[1)] calculated in this work at T = 1423 K.
[2)] experimental data obtained at T = 1423 K after 912 hours of annealing treatments.

### *3.1.2. Elastic and thermal properties*

The elastic constants $C_{ij}$ of the strain tensor were determined by successively imposing distortion matrices $\boldsymbol{D_i}$ [33] on the lattice of $Co_7Mo_6$, according to the expressions

$$\boldsymbol{R} \cdot \boldsymbol{D}_i = \boldsymbol{R}' \qquad (16)$$

$$\boldsymbol{R} = \begin{pmatrix} \frac{\sqrt{3}}{2} & \frac{1}{2} & 0 \\ -\frac{\sqrt{3}}{2} & \frac{1}{2} & 0 \\ 0 & 0 & 1 \end{pmatrix} \qquad (17)$$

where, $\boldsymbol{R}$ is the lattice of μ-phase in a matrix form described by Bravais lattice vectors of a hexagonal crystal with parameters *a* and *c*, $\boldsymbol{R'}$ is the deformed matrix containing distorted lattice vectors.

The variations in total energies $\Delta E(\delta)$ as functions of applied strains for the different types of distortions $\boldsymbol{D_i}$ are plotted in Figure 2.

The calculated elastic constants $C_{ij}$ are listed in Table 2, together with the available theoretical data [27] for comparison. As can be seen from Table 2, $C_{ij}$ satisfy the mechanical stability criterion for hexagonal crystals at zero pressure [28]

$$C_{11} > 0; (C_{11} \cdot C_{33} - 2C_{13}^2 + C_{12} \cdot C_{33}) > 0;\ C_{11} - |C_{12}| > 0;\ C_{44} > 0 \qquad (18)$$

and, as a consequence, the $Co_7Mo_6$ is mechanically stable compound at zero pressure.

**Table 2**
Ground state elastic constants $C_{ij}$ (in GPa) of $Co_7Mo_6$ with theoretical values taken from [27] for comparison.

| Compound | Method | $C_{11}$ | $C_{12}$ | $C_{13}$ | $C_{33}$ | $C_{44}$ | $C_{66}$ |
|---|---|---|---|---|---|---|---|
| $Co_7Mo_6$ | This work | 478.67 | 174.25 | 141.55 | 520.11 | 105.07 | 152.21 |
| | calc. [27 ] | 464 | 191 | 149 | 470 | 103 | 136 |

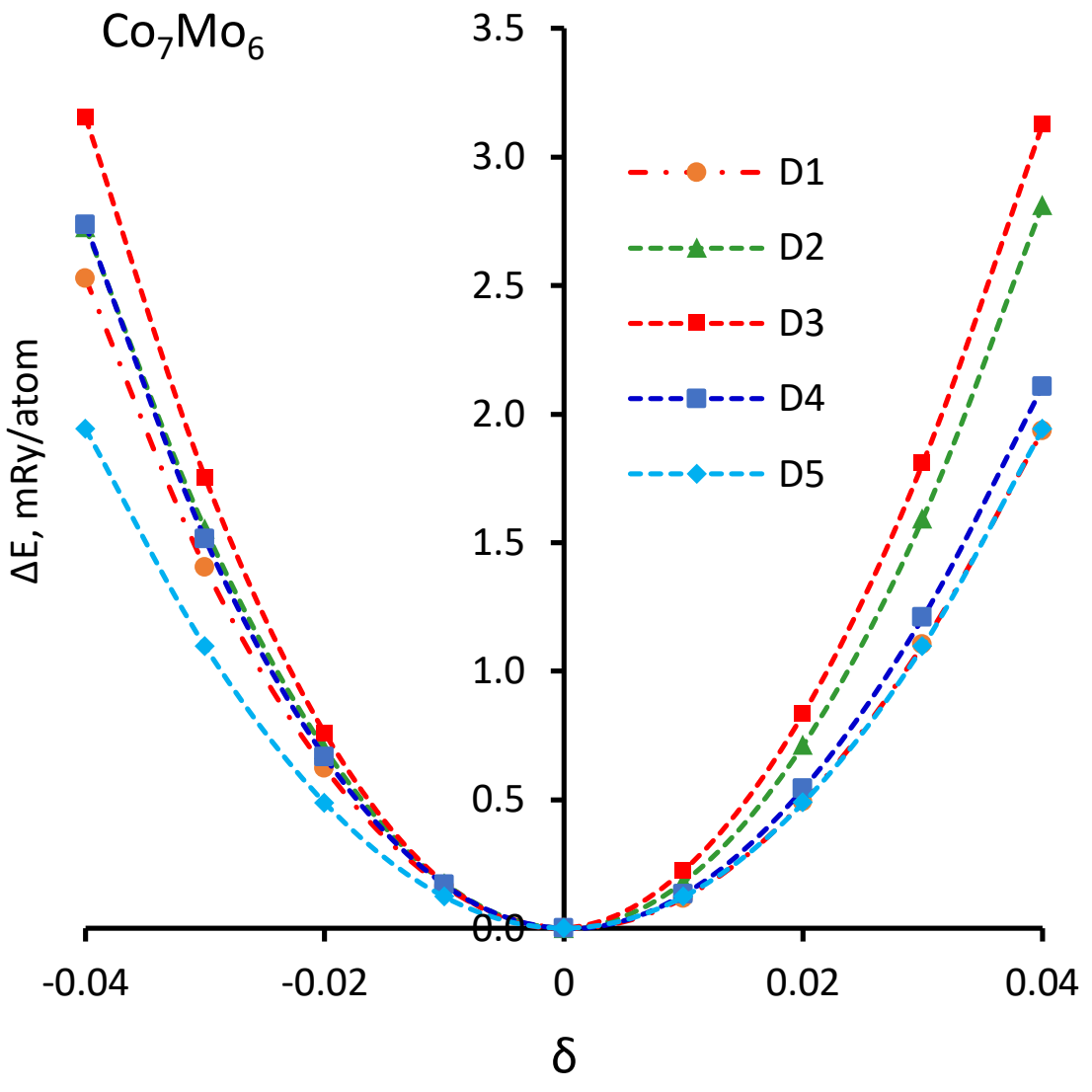

**Figure 2**. The curves of differences in total energies ($\Delta E= E_{tot}(V, \delta_i)-E_{tot}(V_0,0)$, *mRy/atom*) as a functions of applied strains ($\delta$) occurring while imposing the distortion matrices $\boldsymbol{D_1} \div \boldsymbol{D_5}$ on the lattice $Co_7Mo_6$, calculated at T = 0K. The dotted lines are approximations of polynomial functions.

From the $C_{ij}$ obtained for a single crystal, using the Voigt-Reuss-Hill approximation [29], the elasticity parameters for polycrystals were calculated: bulk *B*, shear *G* moduli, Young's modulus *E*, and Poisson's ratio *ν* according [33]. The obtained values presented in Table 3 together with data [27] for comparison.

**Table 3**
The elastic modulus (in GPa) and Poisson's ratio *ν* for $Co_7Mo_6$ calculated at T = 0 K together with theoretical values taken from [27] for comparison.

| Compound | Method | $B_V$ | $B_R$ | $B$ | $G_V$ | $G_R$ | $G$ | $E$ | $\nu$ | $B/G$ |
|---|---|---|---|---|---|---|---|---|---|---|
| $Co_7Mo_6$ | This work | 265.8 | 265.8 | 265.8 | 140.1 | 122.1 | 131.3 | 338.2 | 0.29 | 2.02 |
| | calc. [27] | 264 | 264 | 264 | 129 | 124 | 126 | - | 0.29 | 2.10 |

The values of the *B/G* > 1.75 and ν ≥ 0.26 ratios, according [30, 31], indicate that the $Co_7Mo_6$ polycrystalline aggregate is a ductile material at T = 0K.

**Table 4**
Predicted ground state average ($V_m$), shear ($v_s$) and longitudinal ($v_l$) sound wave velocities (in m/s); Debye temperature $\theta_D$ (in K), Curie temperatures *Tc* (in K) and sound velocities along [001] and [100] crystallographic directions (in m/s) for $Co_7Mo_6$.

| Compound | $v_s$ | $v_l$ | $V_m$ | $\theta_D$ | $T_C$ | [001] | | [100] | | |
|---|---|---|---|---|---|---|---|---|---|---|
| | | | | | | $V_l$ | $V_s$ | $V_l$ | $V_{s1}$ | $V_{s2}$ |
| $Co_7Mo_6$ | 3635 | 6651 | 4054 | 517 | 40 | 7235 | 3252 | 6941 | 3914 | 3252 |

The average ($V_m$), shear ($v_s$), and longitudinal ($v_l$) elastic wave velocities calculated by the formulas [32], predicted Debye $\theta_D$ and Curie temperatures calculated by (9, 12), elastic wave velocities calculated by [33] in the [001] and [100] directions for the $Co_7Mo_6$ μ-phase, are given in Table 4. The Debye temperature is a fundamental parameter of thermodynamics and is found in the equations describing physical processes such as heat capacity, thermal expansion, elastic properties, vibrational entropy, and melting temperature that result from theories involving atomic vibrations. $\theta_D$ is necessary to use the quasi-harmonic theory of Debye - Gruneisen to calculate the thermodynamic properties of $Co_7Mo_6$.

Anisotropic sound waves velocities calculated for $Co_7Mo_6$ are listed in Table 4. The fastest propagation of sound waves was obtained along the [001] direction by longitudinal waves.

For the best knowledge, there are no experimental or theoretical studies on the Curie, Debye temperature and elastic wave velocities for $Co_7Mo_6$. So, the obtained predictions may be useful in further studies of the μ-phase.

### *3.2. Thermodynamic properties at finite temperatures*

*3.2.1. Scheme of the calculations*

In order to investigate the stability of $Co_7Mo_6$ with different lattice parameters and find out whether the thermal expansion paths of this compound is isotropic or not, the approach of searching a thermal expansion path (STEP) [15] was applied.

The layout used in these calculations is shown in Figure 3. The intersection point of the *d0* ÷ *d13* paths (routes) corresponds to the coordinates ($a_0$, $c_0$), these are the optimized parameters of $Co_7Mo_6$ lattice obtained in this work by DFT at T = 0K, which are listed in Table 1. A number of *di* paths are presented here as an example, in Figure 3. The *d0* corresponds to the direction of isotropic expansions of $Co_7Mo_6$, where the *c/a* = 5.359 ratio remains constant. The *d1* route is passing through coordinates of the experimental point obtained at T = 1423 K after 912 hours of annealing treatments [26], as shown in Figure 3 by a blue triangle. Along the *d10* the *a*- parameter of $Co_7Mo_6$ lattice remains constant, as shown in Figure 3.

Since, the volume of the μ-phase, $V(a,c) = (\sqrt{3}/2)\cdot a^2\cdot c$, is a function of lattice parameters *a* and *c*, and in order to avoid dealing with the energy surface *E(a,c)*, the total energies *E(V)* were calculated along the *d0* ÷ *d13* routes, as shown in Figure 4. This approach simplifies the calculations and allows us to reduce the problem to a one-dimensional case and consider the free energy as depending on one variable, the volume. On the one hand, it allows to avoid the differentiation of the total energy *E(a,c)* with respect to variables *a* and *c*, in order to obtain the pressure *P(a,c)* and the other functions, like the bulk modulus *B(a,c)*. And, on the other hand, it allows to use the equation of state (EOS), like the Vinet EOS [34], and apply the quasi-harmonic Debye- Grüneisen model [14] for each paths *di* to simulate the effect of temperature on thermodynamic properties of compounds by calculating the free energies *F(V,T)*, depending only on the volume and temperature, calculated by (1) along each routes. Then, by comparing these free energies between themselves one can find the most energetically favourable thermal expansion route with the lowest energy.

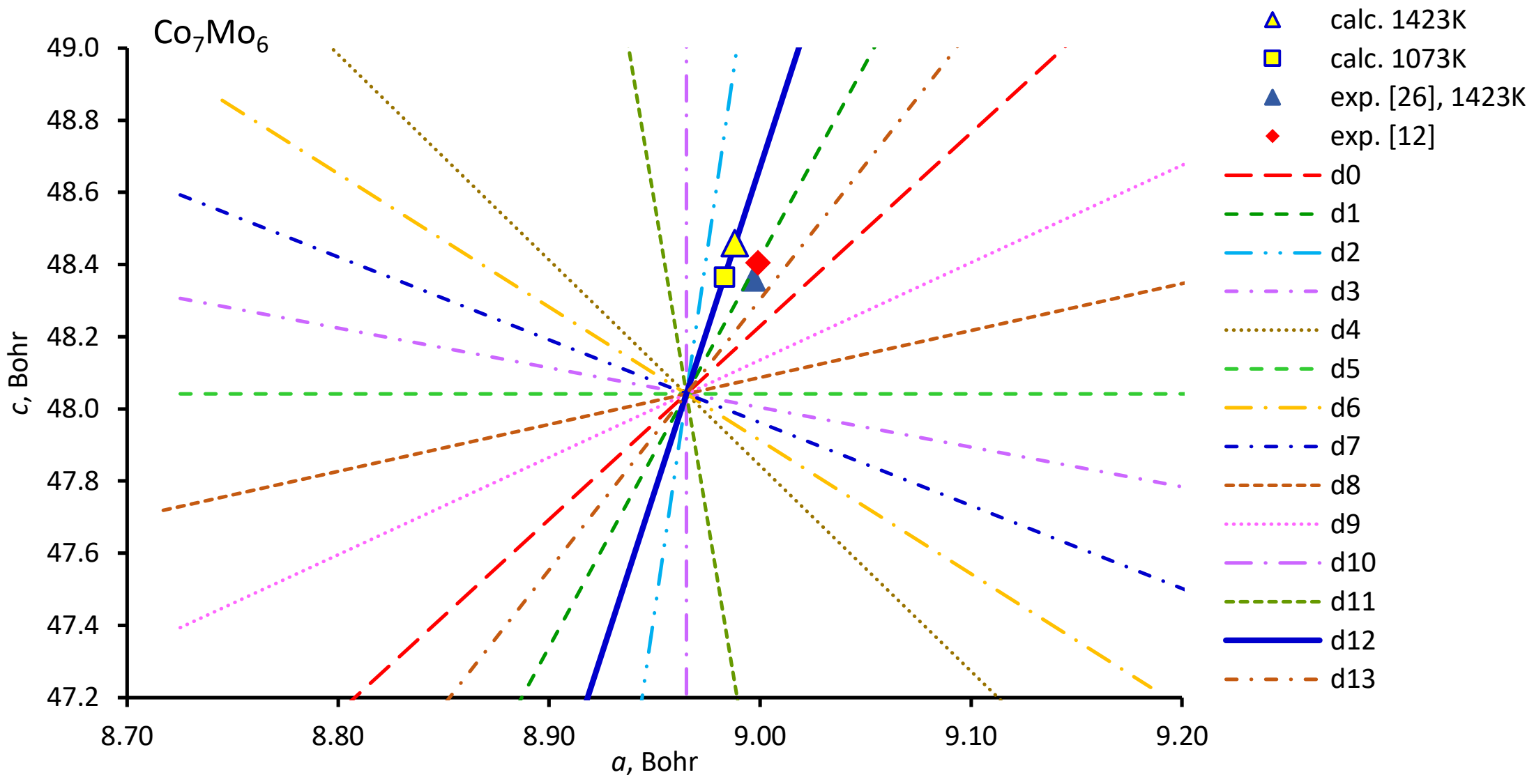

**Figure 3**. The scheme of the calculations of $Co_7Mo_6$ compound. The intersection point of all routes corresponds to the optimized lattice parameters ($a_0$, $c_0$) calculated in this work by DFT at T = 0 K. The red dashed lines *d0* corresponds to the isotropic thermal expansion of $Co_7Mo_6$; the blue solid line *d12* is the calculated thermal expansion path of $Co_7Mo_6$. The calculated in this work at T = 1073 K and T = 1423 K, and experimental lattice parameters of $Co_7Mo_6$ [12], and obtained at T = 1423 K [26] are presented as well.

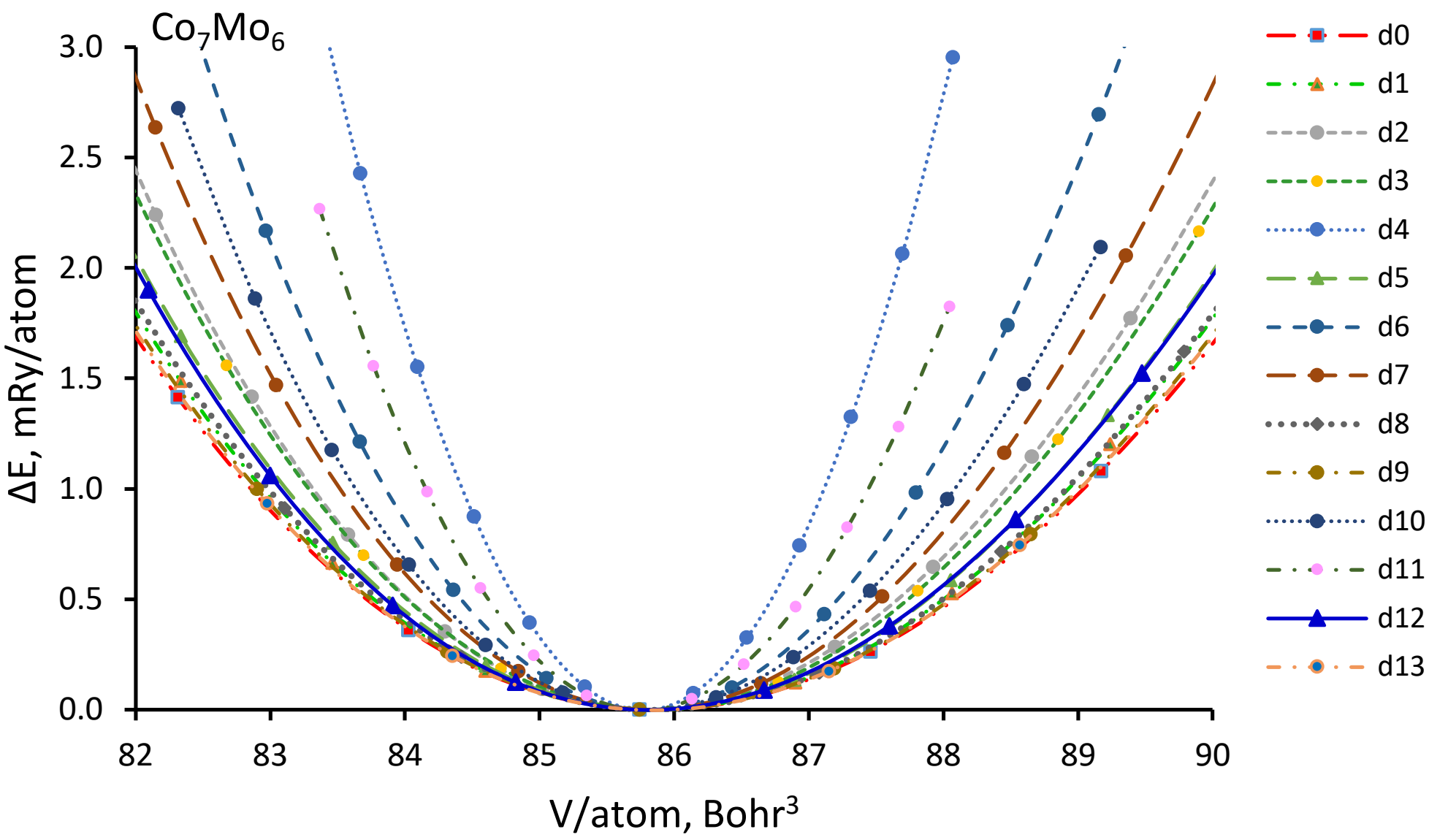

**Figure 4**. Curves of total energies $E_{tot}(V)$ vs volumes obtained by DFT for $Co_7Mo_6$ along the *d0 ÷ d13* routes.

If adding more routes, one can improve an accuracy of the calculations by comparing the energies calculated along these extra routes. In effect, the thermal expansion route may turn out to be curved.

### *3.2.2. Calculations of the Debye temperatures*

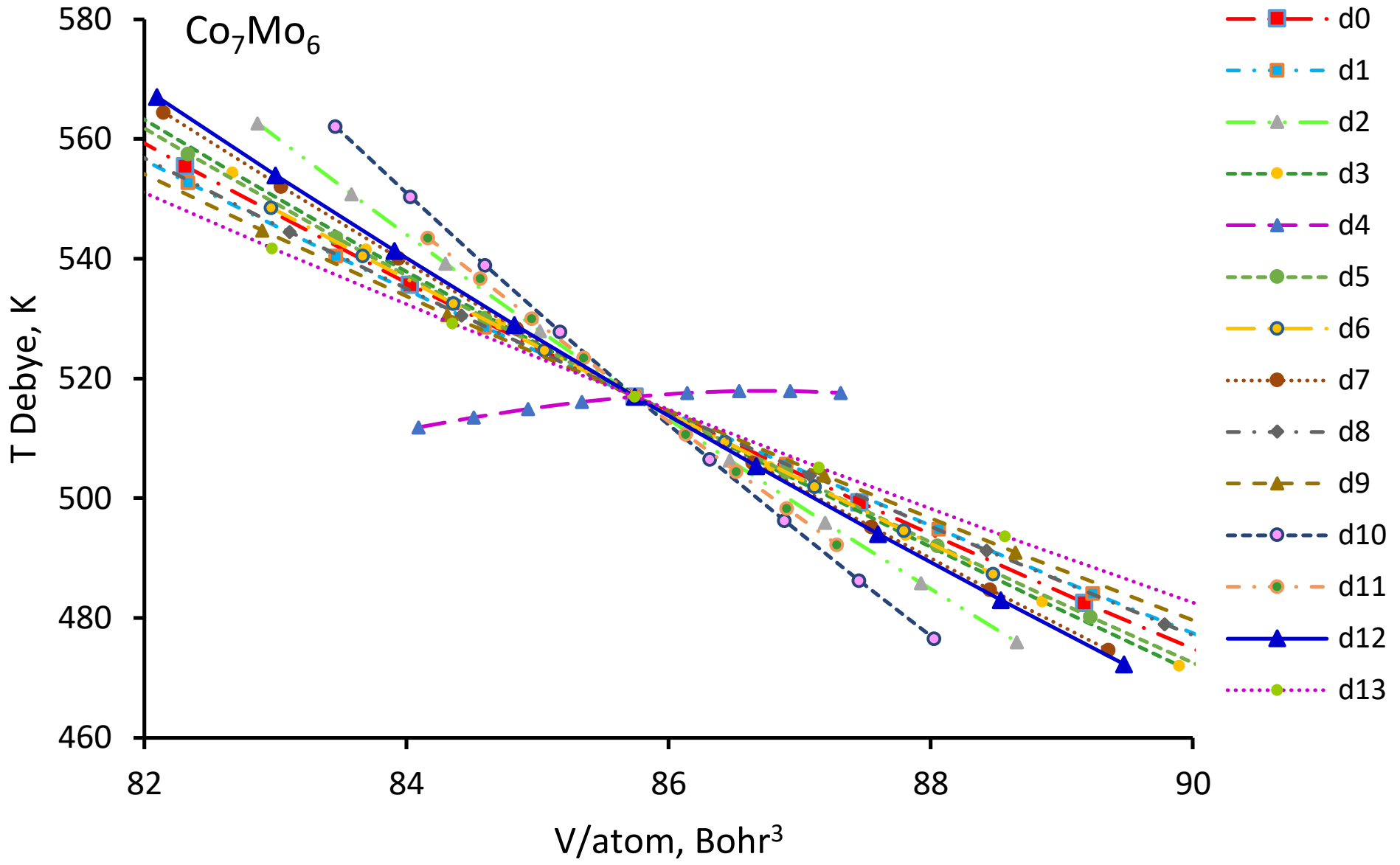

**Figure 5**. Curves of Debye temperatures $\theta_D(V)$ vs volumes calculated for $Co_7Mo_6$ along the *d0 ÷ d13* routes.

To study the vibrational entropy contribution to the free energy of the $Co_7Mo_6$ compound the Debye temperatures $\theta_D(V)$ were calculated for the *d0 ÷ d13* routes by (9) and shown in Figure 5. As follows from the Figure, all the $\theta_D(V)$ curves have negative slopes at the equilibrium volume $V_0$ and at T = 0K, except for the curve calculated along *d4*. The absolute value of the *d4* slope is the smallest among the other, and because that the total energy *E(V)* calculated for *d4* route have the higher energy under expansion and contraction relatively the other routes as follows from Figure 4, therefore, according the Debye–Grüneisen model [14] the contribution to the free energy along the *d4* will be smallest. Along the *d10* route the Debye temperature has the smallest value, but the total energy *E(V)* calculated along the *d10* don't have the smallest value at expansion. And it is difficult to see in advance which route wins this peculiar competition. Thus, all these routes will compete with each other for the most energetically stable thermal expansion path of $Co_7Mo_6$.

### *3.2.3. Accounting for magnetic excitations*

If the magnetic energy is not taken into account, then the thermal expansion, as was shown in [15] using the Laves phase $Fe_2Mo$ as an example, will be isotropic, which contradicts the experimental data. So, in order to investigate the influence of magnetic entropy on the stability of $Co_7Mo_6$ compound the dependences of local magnetic moments, in Bohr magnetons ($\mu_B$), on volumes were obtained along *d0* ÷ *d13* routes and shown in Figure 6.

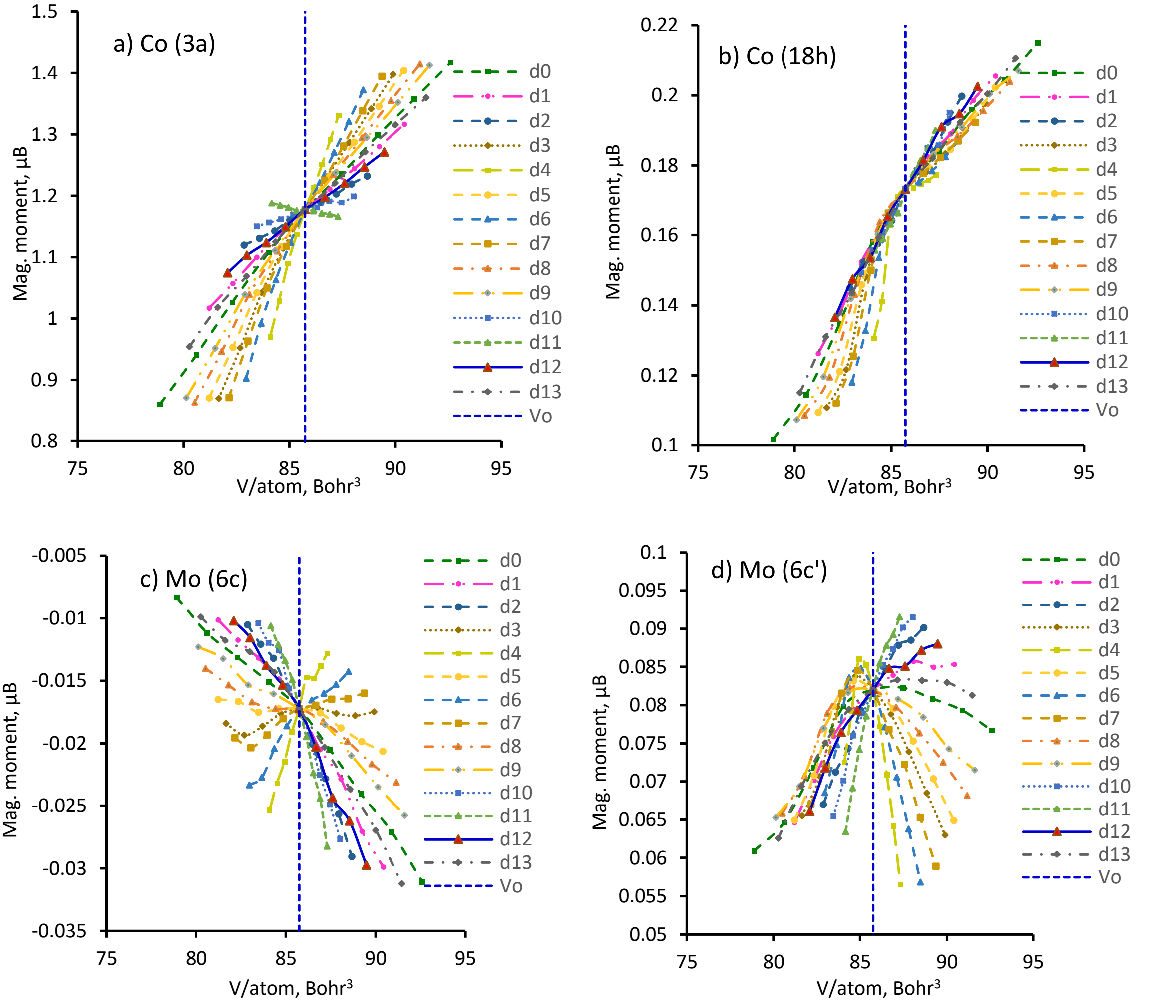

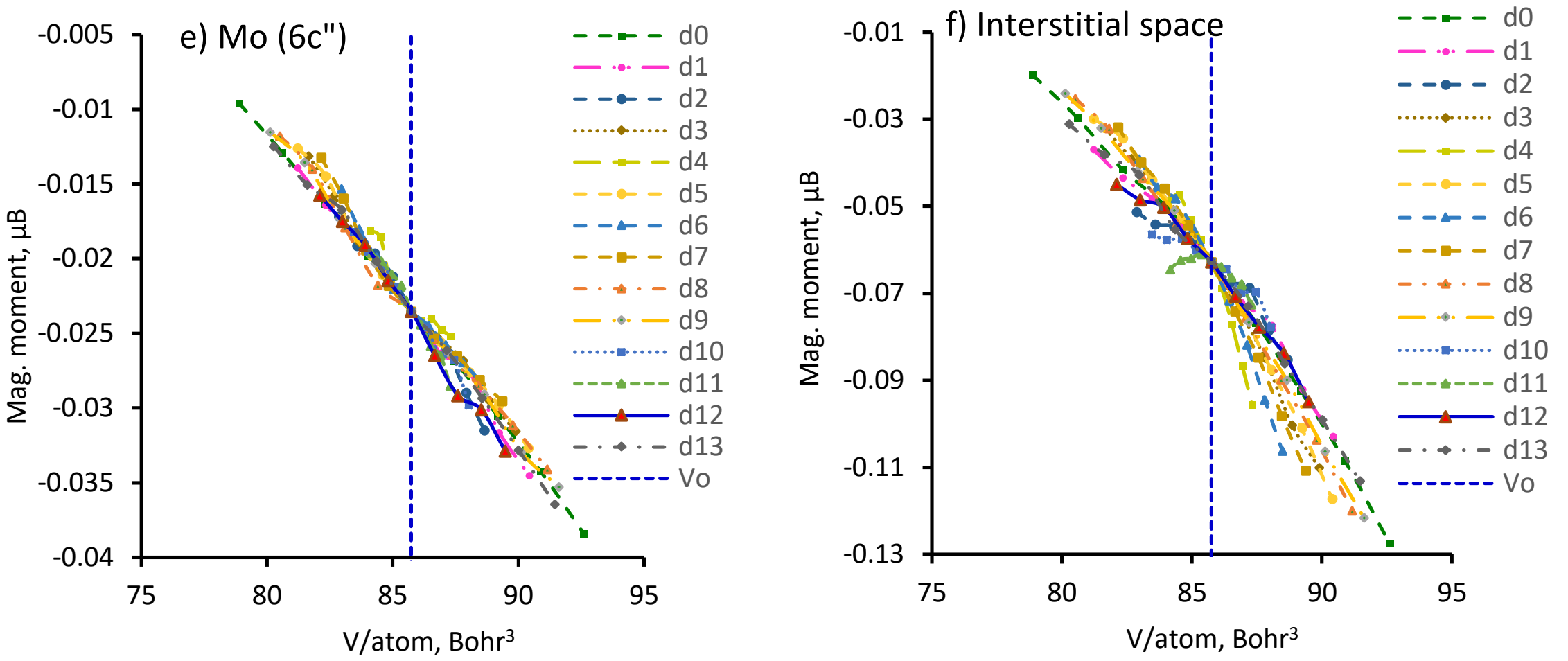

**Figure 6**. Distribution of local magnetic moments ($\mu_B$) on different sublattices and in the interstitial space of $Co_7Mo_6$ μ-phase calculated for *d0* ÷ *d13* routes: a) Co atoms on the first sub-lattice, Co (*3a*); b) Co atoms on the second sub-lattice, Co (*18h*); c) Mo atoms on the third sub-lattice, Mo (*6c*); d) Mo atoms on the fourth sub-lattice, Mo (*6c'*); e) Mo atoms on the fifth sub-lattice, Mo (*6c''*); f) in the interstitial space. $V_0$ – equilibrium volume.

Analyzing the curves of the distribution of magnetic moments over the sublattices obtained along different routes, it was found that at equilibrium volume $V_0$, which is related to T = 0 K, the largest contribution to the magnetic energy of $Co_7Mo_6$ is made by Co atoms located on the *18h* sub-lattice, about 35 %, then follow Co atoms located on the *3a* sub-lattice and the interstitial space, they contribute 28 %, each. The rest comes from Mo atoms accommodated on *6c*, *6c'* and *6c''*, these are 1%, 6% and 2% respectively.

*3.2.4. Helmholtz free energy and its constituents*

The calculations of free energies *F(V,T)* of $Co_7Mo_6$ were carried out by (1) at finite temperatures along the *d0* ÷ *d13* routers shown in Figure 3. The example of free energies calculated along the *d12* route, for T = 100 ÷ 1500K, is shown in Figure 7. Total energy $E_{tot}(V)$ calculated by DFT at T = 0 K along the *d12* route is shown in Figure 7 by the red dashed line together with equilibrium volumes $V_0(T)$ obtained at each temperature as minimum values of the thermodynamic functions and shown by blue dotted line.

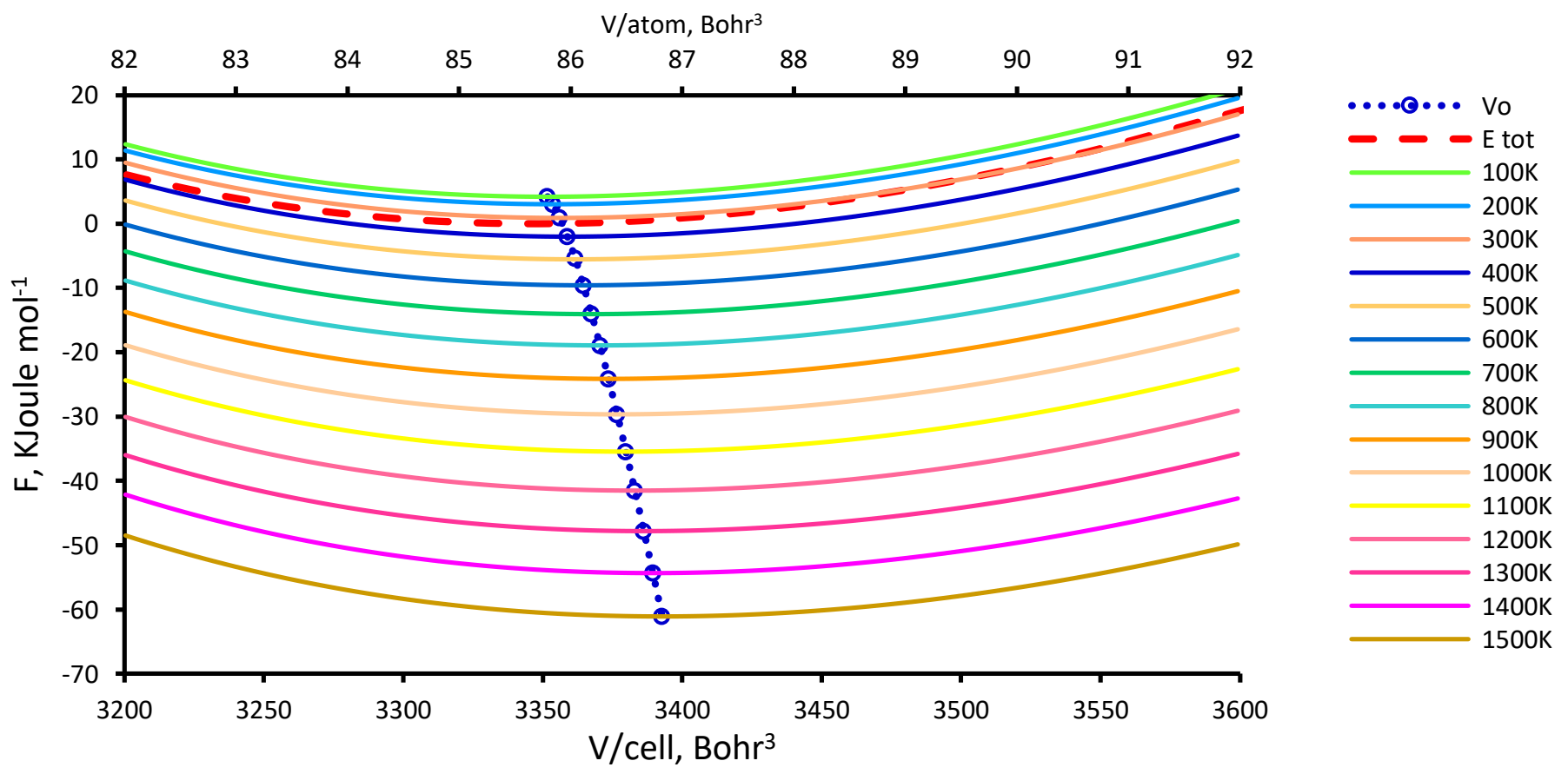

**Figure 7**. Curves of Helmholtz free energy *F(V,T)* calculated versus volumes at finite temperatures for $Co_7Mo_6$ along the *d12* path. The total energy $E_{tot}(V)$ is shown by red dashed line. The volume *V(T)* is shown by blue dotted line, where open circles denote equilibrium volume.

All the energy contributions to the Helmholtz free energy *F(V,T)*: electronic, vibrational and magnetic were calculated by (2 – 14) at the corresponding equilibrium volumes $V_0$.

The graphs of electronic $F_{el}(V_0(T))$, vibrational energies $F_{vib}(V_0(T))$, magnetic entropies, $-T{\cdot}S_{mag}(V_0(T))$, multiplied by -T and the sum of free energies $F(V_0(T))$ calculated by (1-14) vs temperatures for *d0 ÷ d13* routes are shown in Figure 8.

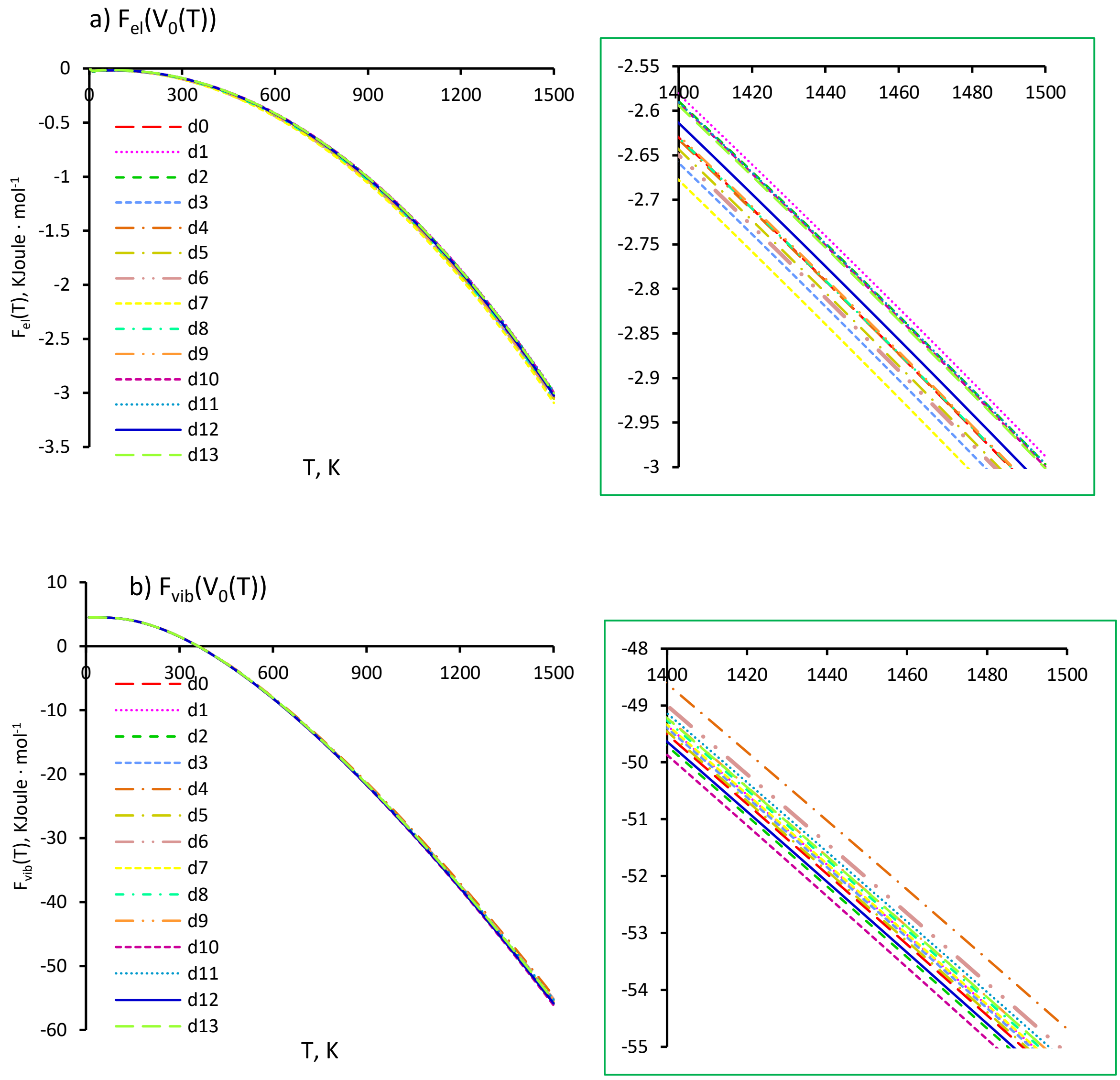

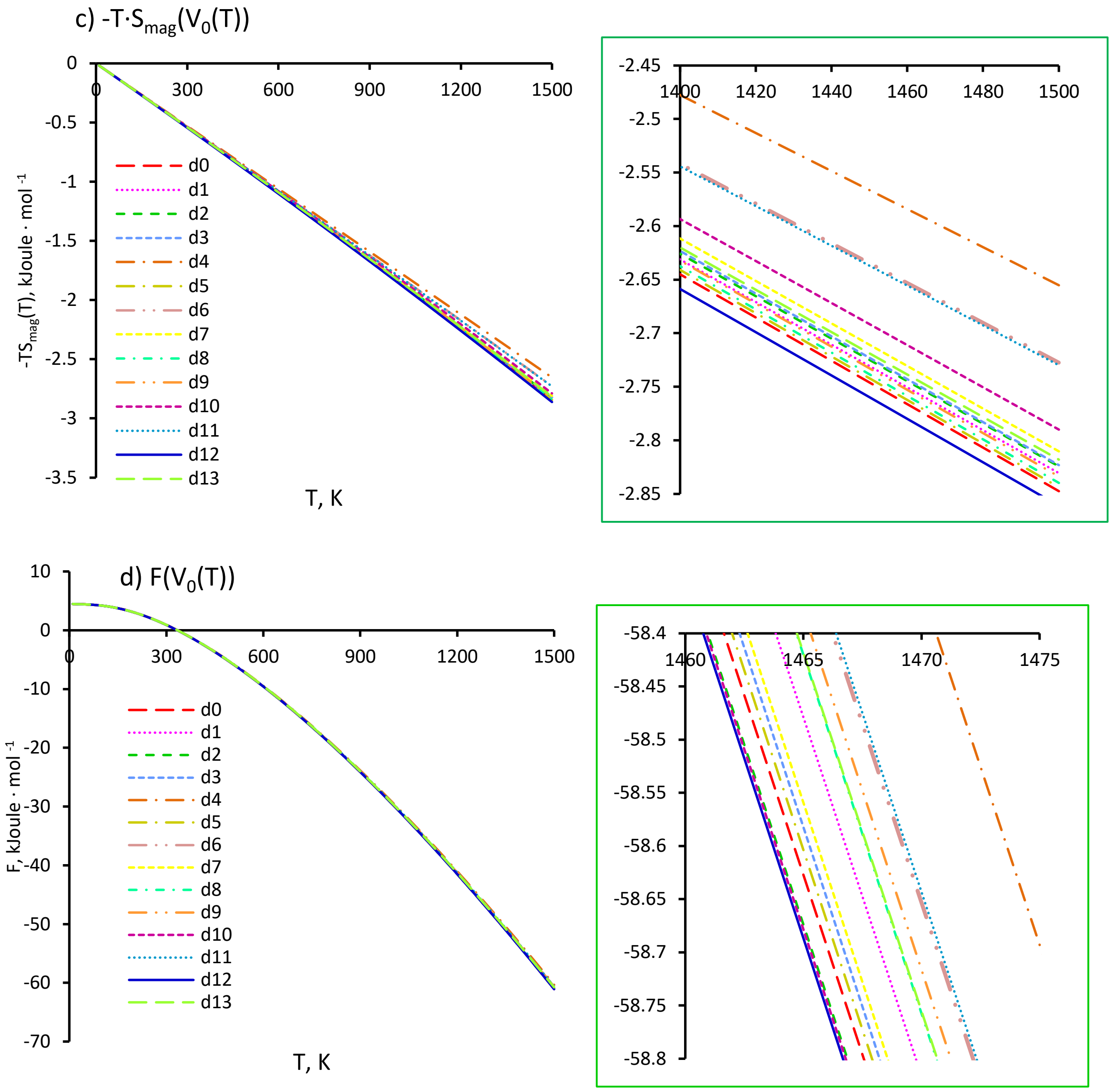

**Figure 8**. The free energy curves *F(T)* of $Co_7Mo_6$ and its energy terms calculated for *d0 ÷ d13* routes. a) Electronic energies, $F_{el}(V_0(T))$; b) Vibrational energies, $F_{vib}(V_0(T))$; c) Magnetic entropies, $-T{\cdot}S_{mag}(V_0(T))$; d) Free energies, $F(V_0(T))$. On the right sides of the graphs, in the inserts, magnified values are shown for convenience.

By analyzing the graphs presented in Figure 8 one can conclude that the electron energy along the *d7* route is the lowest electron energy among others as follows from Figure 8 (a). The *d10* route brings more vibrational energy to the free energy relatively other routes, as shown in Figure 8 (b). The magnetic contribution for the $F(V_0(T))$ benefits from the *d12* route, as presented in Figure 8 (c).

It's essential to note that the configurational entropy $S_{conf}$ (14) in the stoichiometry case, such as $Co_7Mo_6$, gives the same energy contribution for all the routes, and does not affect the comparative analysis. Therefore, by analyzing the free energies $F(V_0(T))$ of $Co_7Mo_6$ shown in Figure 8 (*d*), one can find out that the *d12* route is the most energetically favorable one. Therefore, at heating the lattice parameters of $Co_7Mo_6$ will be increased from ($a_0$, $c_0$), calculated at T = 0K, towards values lying on the *d12* route shown in Figure 3. Thus, this is the pathway of thermal expansion of $Co_7Mo_6$, in accordance with Figure 7.

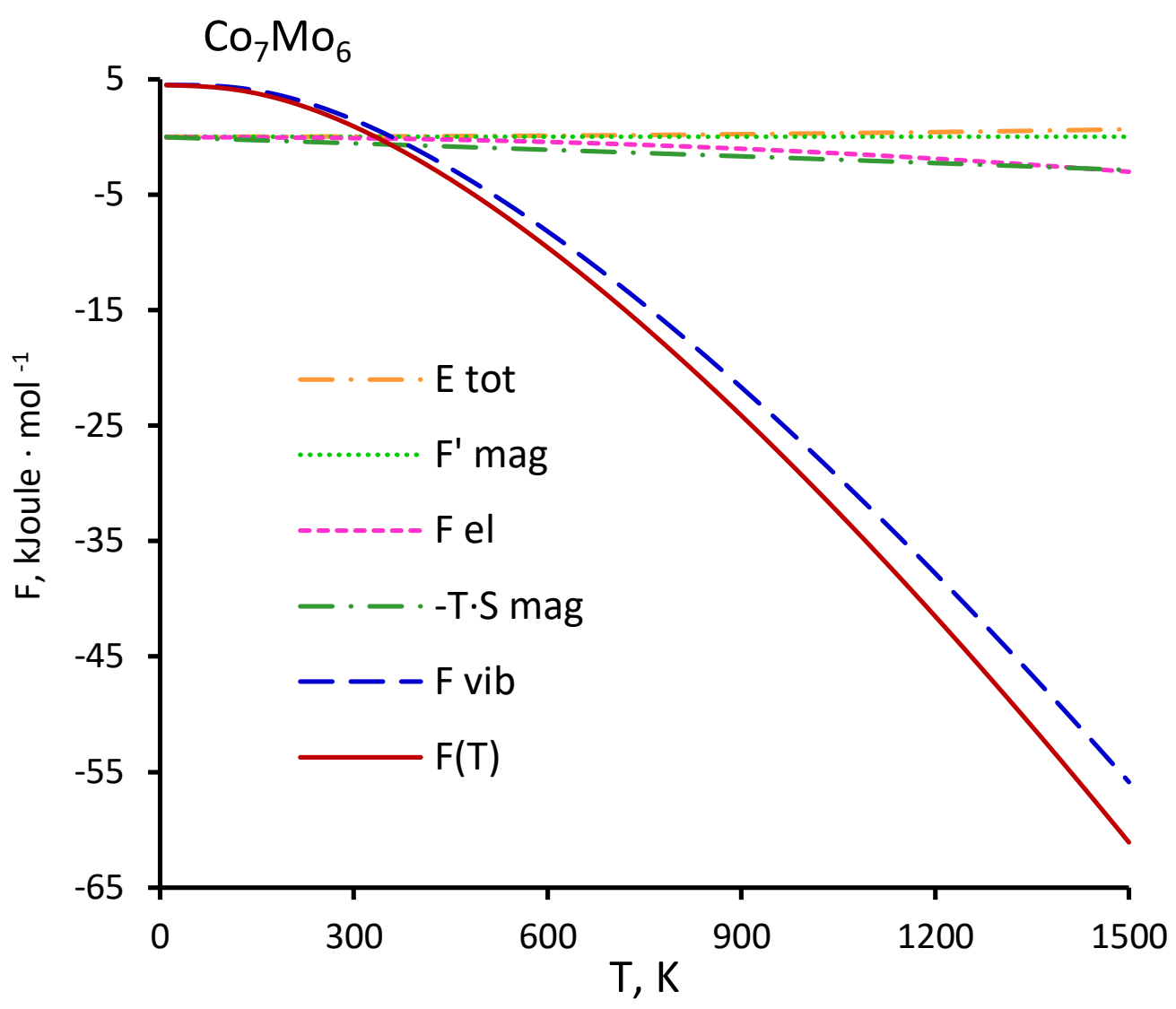

**Figure 9**. The graphs of free energies *F(V₀(T))* for $Co_7Mo_6$ with correspondent terms: total, magnetic, electronic, magnetic entropy and vibrational energies calculated along the *d12* path.

The free energy $F(V_0(T))$ of $Co_7Mo_6$ and its energy terms: total, $E_{tot}(V_0(T))$; magnetic, $F'_{mag}(V_0(T))$; electronic, $F_{el}(V_0(T))$; vibrational, $F_{vib}(V_0(T))$ and magnetic entropy, $-T{\cdot}S_{mag}(V_0(T))$, calculated along the *d12* route are shown in Figure 9.

Analysis shows that at heating the spread of the magnetic energies of $Co_7Mo_6$ in different directions is about 5%, as shown in Figure 8 (c), while the electronic energy is about 2%, as follows from Figure 8 (a). Both the electronic and magnetic energies equally affect the stability of $Co_7Mo_6$. But, the main contributor in the free energy of $Co_7Mo_6$ compound is the vibrational energy. And, the calculation shows that all the energy contributions should be accurately accounted for correct prediction of a thermal expansion path.

It is follow from the obtained free energies of $Co_7Mo_6$ shown in Figure 8 (d) that the other closest stable routes to the *d12* are the *d10, d2* and *d0,* thus these routes lose this peculiar competition.

### *3.2.5. Thermal expansion and heat capacity*

The thermodynamic properties of $Co_7Mo_6$ were obtained along the most energetically favourable thermal route *d12*. The volume expansion *V(T)* of $Co_7Mo_6$ is shown in Figure 10 (a) together with the experimental result [26] obtained at 1423 K after 912 hours of annealing treatments for the sake of comparison. The heat capacity *Cp(T)* of $Co_7Mo_6$ calculated along the route of thermal expansion *d12* is shown in Figure 10 (b) together with the experimental data taken from [5]. The figure also shows the upper and lower error of the experiment, which is given in the work [5]. The heat capacity jump at $T_C$, shown in the graph, is the ferromagnetic/paramagnetic phase transition occurring in $Co_7Mo_6$ upon heating, as the used model [19] predicts. But, since the experimental data on Curie temperatures for $Co_7Mo_6$ is unknown, the search for the exact temperature value is necessary and it is a subject for further study.

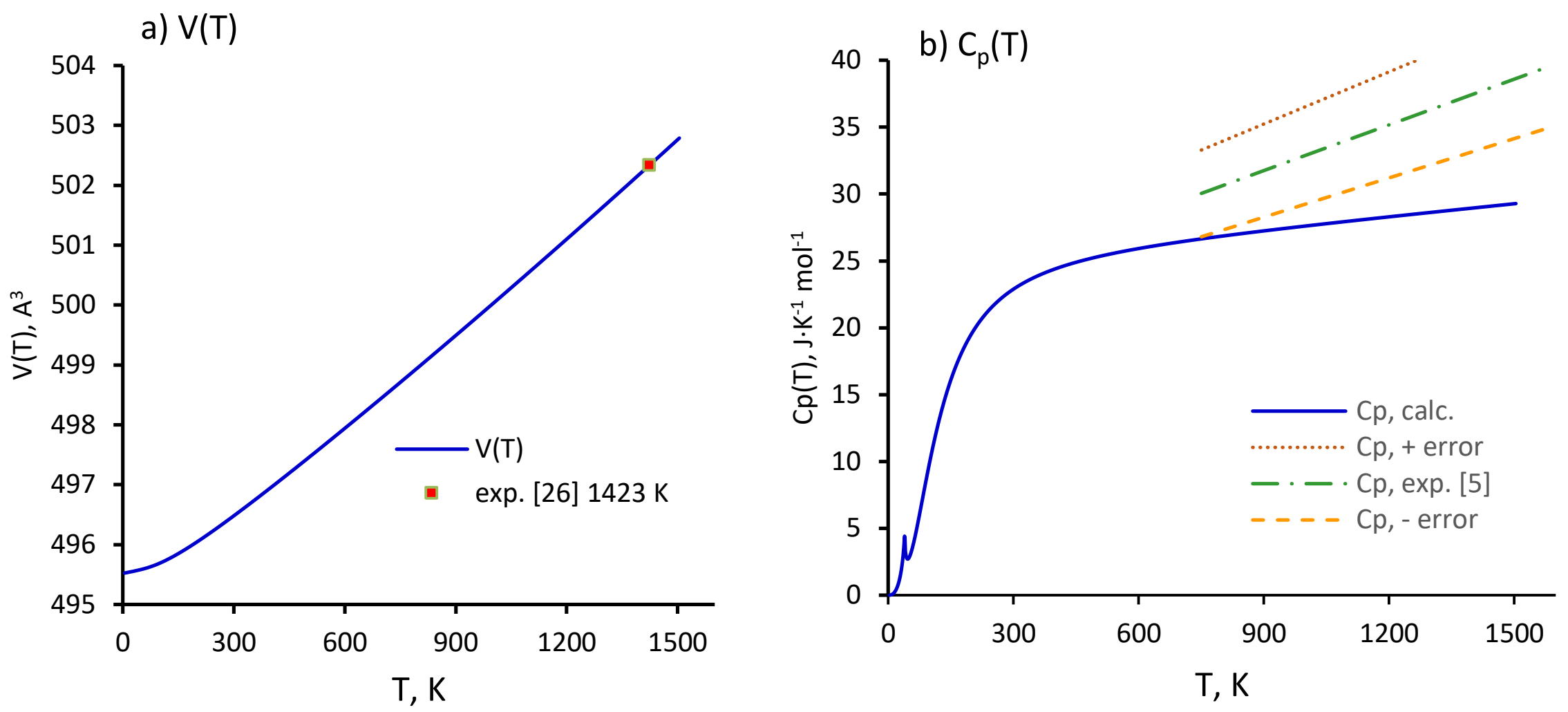

**Figure 10.** a) The volume *V(T)* of $Co_7Mo_6$ calculated along the thermal expansion path is shown in comparison with the experiment data obtained in [26] at T = 1423 K after annealing of 912 hours; b) the heat capacity *Cp(T)* of $Co_7Mo_6$ calculated along the route of thermal expansion is presented together with the experimental data taken from [5] and the scatter of experimental data, which provided in the work [5], is shown as "+ error" and "- error".

### *3.3. Discussion*

Thus, DFT calculations with quasi-harmonic Debye - Grüneisen approximation and by employing the approach of searching a thermal expansion path (STEP) of a compound [15] which compares free energies calculated along different routes of thermal expansion, can correctly predict the thermal expansion of $Co_7Mo_6$ compound. The lattice parameters of $Co_7Mo_6$ calculated in this work at T = 1423 K are in a good agreement with the experimental data reported in [12, 26] as shown in Figure 3, Figure 10 (a) and listed in Table 1. The heat capacity calculated in this work and the experimentaly estimated data reported in work [5], as shown in Figure 10 (b), are in satisfactory agreement at T = 800 K, at the higher temperatures the calculated heat capacity increases slowly then the estimated values [5].

The results of this work show that if we neglect the electronic and magnetic energies, then the *d10* route, along which the lattice parameter *a* remains constant, will be the most energetically favourable, as shown in Figure 8 (b) and presented in Figure 5 by the most lower Debye temperature along the *d10*. But the experimental data on the lattice parameters obtained in [26] after annealing at T = 1423 K, and in [12], show that this is not the case. The calculated *d12* path of thermal expansion is passing close to the experimental data [12, 26] as shown in Figure 3. Therefore, the electronic and magnetic entropies have an essential effect on the stability of this compound, and in no case should they be neglected.

The phonon calculations of the vibrational energy and their comparison with energies calculated in this work using the Debye-Grüneisen model, which theoretically should match each other, may be a valuable argument for checking the results obtained in this work.

Similar calculations were carried out in work [35], where the direction of thermal expansion of the $Fe_2Mo$ Laves phase calculated using the phonon spectrum was compared with the direction calculated using the Debye-Gruneisen approximation. Comparative analysis showed that these two methods completely coincide with each other and can be used in similar STEP calculations.

## 4. Conclusions

The finite-temperature DFT calculations have been carried out to calculate electronic, vibrational and magnetic free energy contributions for the $Co_7Mo_6$ μ-phase. By employing the approach of searching a thermal expansion path of compounds the thermal expansion of $Co_7Mo_6$ is correctly calculated in accordance with the experimental data. The nature of the thermal expansion is not isotropic. The elastic constants, bulk modulus, elastic sound velocities, Curie and Debye temperatures were calculated at ground state. The heat capacity calculated along the thermal expansion path of $Co_7Mo_6$ is in satisfactory agreement with the experimental data. Thus, the approach of searching a thermal expansion path of compounds may be used to study thermodynamic properties of μ-phase. The work shows that the vibrational energy is the main factor influencing on the stability of $Co_7Mo_6$, but the electronic and magnetic entropies should be accurately accounted for correct thermodynamic description of the μ-phase.

## Acknowledgments

The research was financially supported by the Russian Foundation for Basic Research as a part of scientific project № 19-03-00530.